\documentclass{aa}
\usepackage[dvips]{graphicx}
\usepackage{txfonts}
\def\simless{\mathbin{\lower 3pt\hbox
{$\rlap{\raise 5pt\hbox{$\char'074$}}\mathchar"7218$}}}   
\def\simmore{\mathbin{\lower 3pt\hbox
{$\rlap{\raise 5pt\hbox{$\char'076$}}\mathchar"7218$}}}   
\def\Msun{{\rm M}_\odot}                                       
\newcommand{\be}{\begin{equation}}
\newcommand{\ee}{\end{equation}}

\begin{document}
\title{Neutrino-cooled accretion and GRB variability}
\titlerunning{Neutrino-cooled accretion and GRB variability}
\author{Dimitrios Giannios}

\institute{Max Planck Institute for Astrophysics, Box 1317, D-85741 Garching, Germany}

\offprints{giannios@mpa-garching.mpg.de}
\date{Received / Accepted}

\abstract
{For accretion rates ${\dot M}\sim 0.1$ $\Msun$/s to a few solar mass black
hole the inner part of the disk is expected to make a transition from advection dominance to 
neutrino cooling. This transition is characterized by sharp changes of the disk properties.
I argue here that during this transition, a modest increase of the accretion rate leads to
powerful enhancement of the Poynting luminosity of the GRB flow and decrease of its
baryon loading. These changes of the characteristics of the GRB flow translate
into changing gamma-ray spectra from the photosphere of the flow. 
The photospheric interpretation of the GRB emission explains the observed
narrowing of GRB pulses with increasing photon energy and the 
luminosity-spectral peak relation within and among bursts.

\keywords{Gamma rays: bursts -- Accretion, accretion disks}}

\maketitle

\section{Introduction} 
\label{intro}

The commonly assumed model for the central engine of gamma-ray bursts
(hereafter GRBs) consists of a compact object, most likely a black hole, 
surrounded by  a massive accretion disk. This configuration results
naturally from the collapse of the core of a fast rotating, massive star
(Woosley 1993; MacFadyen \& Woosley 1999) or the coalescence of a neutron
star-neutron star or a neutron star-black hole binary (for simulations
see Ruffert et al. 1997). 

The accretion rates needed to power a GRB are in the range ${\dot M}\sim
0.01-10$ $\Msun$/s. Recently, much theoretical work has been done to understand 
the microphysics and the structure of the disk at this very high
accretion-rate regime (e.g., Chen \& Beloborodov 2007; hereafter CB07). These 
studies have shown that while for accretion rates ${\dot M}\ll 0.1\Msun$/s the 
disk is advection dominated, when  ${\dot M}\sim 0.1\Msun$/s it makes
a sharp transition to efficient neutrino cooling. This transition results to 
a thiner, much denser and neutron rich disk. 
 
Here I show that, for reasonable scalings of the magnetic field strength with
the properties of the inner disk, the advection dominance-neutrino cooling
transition results in large changes in the Poynting flux output in the GRB flow. 
During this transition, a moderate increase of the accretion
rate is accompanied by large increase of the  Poynting luminosity and decrease of the
baryon loading of the GRB flow. This leads to powerful and ``clean'' ejections of material. 
The photospheric emission from these ejections explains the observed
narrowing of GRB pulses with increasing photon energy (Fenimore et al. 1995)
and the luminosity-spectral peak relation within and among bursts (Liang 
et al. 2004).

\section{Disk transition to efficient neutrino cooling}

In accretion powered GRB models the outflow responsible for the GRB
is launched in the polar region of the black-hole-disk system. 
This can be done by neutrino-antineutrino annihilation and/or
MHD mechanisms of energy extraction. In either case, the power
output in the outflow critically depends on the physical 
properties of the inner part of the accretion disk. In this section, I focus
on the disk properties around the transition from
advection dominance to neutrino cooling. The implications
of this transition on the energy output to the GRB flow are
the topic of the next section.

Recent studies have explored the structure of accretion disks 
that surround a black hole of a few solar masses for accretion
rates ${\dot M}\sim 0.01-10$ $\Msun$/s. Most of these studies focus on 1-D 
``$\alpha$''-disk models (where $\alpha$
relates the viscous stress to the pressure in the disk; Shakura \& Sunyaev
1973) and put emphasis on the treatment of the microphysics of the disks connected
to the neutrino emission and opacity, nuclear composition and electron degeneracy
(Di Matteo et al. 2002; Korhi \& Mineshige 2002; Kohri et al. 2005; CB07; 
Kawanaka \& Mineshige 2007; hereafter KM07) and on general relativistic 
effects on the hydrodynamics (Popham et al. 1999; Pruet at al. 2003; CB07). 

These studies have shown that for ${\dot M}\simless 0.1$ $\Msun$/s and
viscosity parameter $\alpha\sim 0.1$ the disk is advection dominated since   
the large densities do not allow for photons to escape. The temperature at
the inner parts of the disk is $T\simmore 1$ MeV and the density 
$\rho \sim 10^8$ gr/cm$^3$ which results in a disk filled with mildly degenerate pairs. 
In this regime of temperatures and densities the
nucleons have dissociated and the disk consists of free protons and
neutrons of roughly equal number. The pressure in the disk is:
  $P= P_{\gamma,e^{\pm}}+P_{\rm b}$. 
The first term accounts for the pressure coming from radiation {\it and} pairs
and the second for that of the baryons. In the advection dominated regime the
pressure is dominated by the ``light particle'' contribution 
(i.e. the first term in the last expression).  

For accretion rates $\dot{M}\sim 0.1$ $\Msun$/s, a rather sharp
transition takes place in the inner parts of the disk. During this
transition, the mean electron energy is high enough for electron capture by
protons to be possible: $e^-+p\to n+\nu$.  As a result, the disk
becomes neutron rich, enters a  phase of efficient neutrino cooling and
becomes thinner. The baryon density of the disk increases dramatically and
the total pressure is dominated by the baryon pressure. After the transition
is completed the neutron-to-proton ratio in the disk is $\sim 10$.
Hereafter, I refer to this transition as ``neutronization'' transition.

The neutronization transition takes place at an approximately constant disk temperature $T\approx$
several$\times 10^{10}$ K and is completed for a moderate increase of the
accretion rate by a factor of $\approx 2-3$. During the transition the baryon
density increases by $\approx 1.5$ orders of magnitude and the disk pressure
by a factor of several (see CB07; KM07).

\subsection{Scalings of the disk properties with ${\dot M}$}

Although the numbers quoted in the previous section hold quite generally,
the range of accretion rates for which the neutronization transition 
takes place depends on the $\alpha$ viscosity parameter and on the spin of the black hole. For more
quantitative  statements to be made, I extract some 
physical quantities of the disk before and after transition from Figs. 13-15
of CB07 for disk viscosity $\alpha=0.1$ and
spin parameter of the black hole $a=0.95$. I focus at a fixed radius close to
the inner edge of the disk (for convenience, I choose $r=6GM/c^2$). 
The quantities before and after the transition are marked with the superscripts
``A'' and ``N'' and stand for Advection dominance  and Neutrino cooling
respectively. At ${\dot M^{\rm A}}=0.03$ $\Msun$/s, the density
of the disk is $\rho^{\rm A}\simeq 3\cdot 10^{9}$gr/cm$^3$ and has similar number of
protons and neutrons, while at ${\dot M^{\rm N}}=0.07$ $\Msun$/s, the density is $\rho^{\rm N}\simeq
9\cdot 10^{10}$gr/cm$^3$ and the neutron-to-proton ratio is $\sim10$. 
The temperature remains approximately constant for
this range of accretion rates at $T\simeq 5\cdot 10^{10}$ K.
A factor of ${\dot M^{\rm N}}/{\dot M^{\rm A}}\simeq 2.3$ increase in the accretion rate
in this specific example leads to the transition from advection dominance to neutrino cooling.     

Around the transition the (mildly degenerate) pairs contribute a factor
of $\sim 2$ more to the pressure w.r.t. radiation. The total pressure is:  
$P=P_{\gamma,e^{\pm}}+P_{\rm b}\approx a_{\rm r}T^4+\rho k_{\rm B} T/m_{\rm
  p}$, where $a_{\rm r}$ and $k_{\rm B}$ are the radiation and Boltzmann 
constants respectively (Beloborodov 2003; CB07).    
Using the last expression, the disk pressure before
the transition is found: $P^{\rm A}\simeq 6\cdot 10^{28}$ erg/cm$^3$; 
dominated by the contribution of light particles as
expected for an advection dominated disk. At the higher accretion rate ${\dot
M}^{\rm N}$, one finds for the pressure of the disk $P^{\rm N}\simeq
4\cdot 10^{29}$ erg/cm$^3$. Now the disk is baryon pressure supported.  
  
From the previous exercise one gets {\it indicative} scalings
for the dependence of quantities in the disk as a function of ${\dot M}$
{\it during} the neutronization transition: $\rho\propto {\dot M}^{4}$ and $P\propto {\dot
M}^{2.3}$.  Doubling of the accretion rate during
the transition leads to a factor of $\sim 16$ and $\sim 5$ increase 
of the density and pressure of the disk respectively. 

Similar estimates for the dependence of the disk density and pressure on the 
accretion rate can be done when the inner disk is in the advection dominance 
and neutrino cooling regime but fairly close to the transition. In these
regimes, I estimate that $\rho \propto P \propto  {\dot M}$ (see, for example  
Figs. 1-3 in KM07). 

Does this sharp change of the disk properties associated with the
neutronization transition affect the rate of energy release in the polar region of
the disk where the GRB flow is expected to form? The answer depends on the mechanism
responsible for the energy release.

\section{Changes in the GRB flow from the neutronization transition}

Gravitational energy released by the accretion of matter to the black hole
can be tapped by neutrino-antineutrino annihilation or via MHD mechanisms 
and power the outflow responsible for the GRB. 
We consider both of these energy extraction mechanisms in turn.

The neutrino luminosity of the disk just after the neutronization transition is of the order of 
$L_{\nu}\sim 10^{52}$ erg/s and consists of neutrinos and antineutrinos of all
flavors. The fraction of these neutrinos that annihilate and power the GRB flow
depends on their spatial emission distribution which, in turn, depends
critically on the disk microphysics.  For ${\dot M}\sim 0.1 \Msun/$s, this fraction 
is of the order of $\sim 10^{-3}$ (Liu et al. 2007), 
powering an outflow of $L_{\nu{\bar \nu}}\sim 10^{49}$ erg/s; most 
likely too weak to explain a cosmological GRB. The efficiency of the
neutrino-antineutrino annihilation mechanism can be much  
higher for accretion rates ${\dot M}\simmore$ $1\Msun/$s (e.g., Liu et al. 2007;
Birkl et al. 2007) which are not considered here.

The second possibility is that energy is extracted by strong magnetic
fields that thread  the inner part of the disk (Blandford \& Payne 1982) or
the rotating black hole (Blandford \& Znajek 1977) launching
a Poynting-flux dominated flow. The Blandford-Znajek power output can be estimated to
be (e.g. Popham et al. 1999)
\be
L_{\rm BJ}\approx 10^{50}a^2 B_{15}^2M_3^2 \quad \rm erg/s,
\ee      
where $B=10^{15}B_{15}$ Gauss and $M=3M_3\Msun$. taking into account that
magnetic fields of similar strength are expected to thread the inner parts of the disk, 
the Poynting luminosity output from the disk is rather higher than $L_{\rm
 BJ}$ because of the larger effective surface of the disk (Livio et al. 1999). In conclusion,  
magnetic field strengths in the inner disk of the order of $B\sim 10^{15}$
erg/s are likely sufficient to power a GRB via MHD mechanisms of energy
extraction. 

\subsection{Luminosity and baryon loading of the GRB flow as functions of ${\dot M}$}

In this section, I estimate the Poynting luminosity of the GRB flow for 
different assumptions on the magnetic field-disk coupling. The
mass flux in the GRB flow is harder to constrain since it depends on 
the disk structure and the magnetic field geometry on the
disk's surface. During the neutronization transition, the disk becomes 
thinner and, hence, more bound gravitationally. One can thus expect
that a smaller fraction of ${\dot M}$  is injected in the outflow.  
Here, I make the, rather conservative, assumption that throughout the 
transition, the mass flux in the outflow is a fixed fraction of accretion rate
${\dot M}$.     

How is the magnetic field strength related to the properties of the disk?
The magneto-rotational instability (hereafter MRI; see Balbus \& Hawley 1998 for a review) 
can amplify magnetic field with energy
density up to a fraction $\epsilon$ of the pressure in the disk.  This
provides an estimate for the magnetic field: $B_{\rm MRI}^2=8\pi \epsilon P$. This scaling leads
to magnetic field strength of the order of $\sim 10^{15}$ Gauss for the fiducial
values of the pressure presented in the previous Sect. and for $\epsilon\simeq
0.2$. 

The Poynting luminosity scales as $L_{\rm p}\propto B_{\rm MRI}^2\propto P\propto \dot{M}^{2.3}$ 
with the accretion rate during the neutronization transition (see previous
Sect.). This leads to a rather large increase
of the luminosity of the GRB flow by a factor of $\sim 7$ for a moderate
increase of the accretion rate by a factor of $\simeq 2.3$. Furthermore,
if we assume that a fixed fraction of the accreting gas is channeled to the
outflow, then the baryon loading of the Poynting-flux dominated flow
scales as $\eta\propto L_{\rm p}/\dot{M}\propto {\dot M}^{1.3}$. This means that
during the transition the outflow becomes ``cleaner'' decreasing its 
baryon loading by a factor of $\sim 3$.

The disk can support large-scale fields more powerful that those generated by MRI. 
These fields may have been advected with
the matter during the core collapse of the star (or the binary coalescence) or are
captured by the disk in the form a magnetic islands and brought in the 
inner parts of the disk (Spruit \& Uzdensky 2005). These large
scale fields can arguably provide much more promising conditions to launch
a large scale jet. 

Stehle \& Spruit (2001) have shown that a disk threaded by a large scale field becomes violently
unstable once the radial tension force of the field contributes substantially against 
gravity. This instability is suppressed if the radial tension force is a
faction $\delta \sim$ a few \% of the gravitational attraction. Large-scale
magnetic fields with strength: $B_{\rm LS}^2
=\delta 8\pi \rho c_s v_k\propto (\rho P)^{1/2}$ can be supported over the duration of a GRB for
$\delta \sim$ a few \%. In the last expression $c_s=\sqrt{P/\rho}$ stands
for the sound speed and $v_k$ is the Keplerian velocity at the inner boundary.

The last estimate suggests that  large scale field strong enough to power a
GRB can be supported by the disk. The output Poynting
luminosity scales, in this case , as  $L_{\rm p}\propto B_{\rm LS}^2\propto (\rho P)^{1/2}$. During the
neutronization transition, the Poynting luminosity increases steeply as a
function of the accretion rate: $L_{\rm p} \propto (\rho P)^{1/2} \propto {\dot
M}^{3.2}$. This translates to a factor of $\sim
15$ increase of the luminosity of the jet for a modest increase by $\sim 2.3$
of the accretion rate. Assuming that the rate of ejection of material 
in the GRB flow is proportional to the mass accretion rate, the baryon 
loading of the flow is found to decrease by a factor of $\sim 6$ during 
the transition (since $\eta \propto L_{\rm p}/\dot{M}\propto {\dot M}^{2.2}$).

Before and after the transition the disk is advection dominated and 
neutrino cooled respectively. When the disk is in either of these regimes 
the disk density and pressure scale roughly linearly with the
accretion rate (at least for accretion rates fairly close to the neutronization transition; 
see previous Sect.), leading to $L_{\rm p}\propto  {\dot M}$ and $\eta \sim$ constant. 
The Poynting luminosity and the baryon loading of the GRB flow
around the neutronization transition are summarized by Fig.~1.

\begin{figure}
\resizebox{\hsize}{!}{\includegraphics[angle=270]{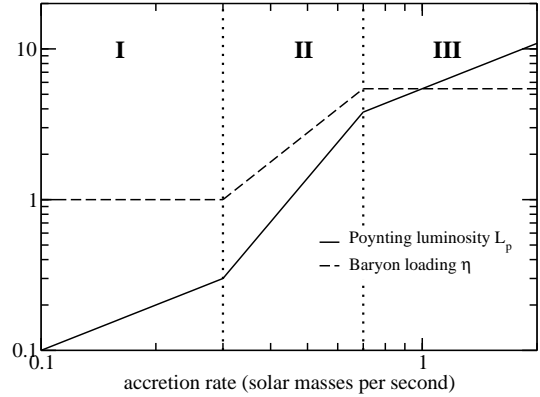}}
\caption[]
{Poynting luminosity and baryon loading (both in arbitrary units)
of the GRB flow around the neutronization transition of the 
inner disk. In regions marked with I and III the inner disk is advection 
and neutrino cooling dominated respectively. In region II, the neutronization transition
takes place. During the transition, the Poynting luminosity increases steeply 
with the accretion rate while the baryon loading of the flow is reduced (i.e.
$\eta$ increases). 
\label{fig1}}
\end{figure}

Although the Poynting flux output depends on  
assumptions on the scaling of the magnetic field with the disk properties,  
the neutronization transition generally leads to steep increase of the Poynting 
luminosity as function of the accretion rate and to a ``cleaner'' (i.e. less baryon loaded)
flow. Observational implications of the transition are discussed in the next section.

\section{Connection to observations}
             
The mechanism I discuss here operates for accretion rates around the neutronization transition of the
inner disk and provides the means by which modest variations in
the accretion rate give magnified variability in the Poynting flux output and baryon
loading of the GRB flow. Since the transition  takes place 
at ${\dot M}\sim 0.1$ $\Msun$/s which is close to the accretion rates 
expected for the collapsar model (MacFadyen \& Woosley 1999), it is
particularly relevant for that model. 
To connect the flow variability to the observed properties 
of the prompt emission, one has to assume a model for the prompt emission.
Here we discuss internal shock and photospheric models.

Episodes of rapid increase of the luminosity of the flow can be viewed as the  
ejection of a distinct shells of material. These shells 
can collide with each other further out in the flow leading to internal shocks 
that power the prompt GRB emission (Rees \& M\'esz\'aros 1994).
For the internal shocks  to be efficient in dissipating energy, there must be a substantial variation of the 
baryon loading  among shells.
This may be achieved, in the context of the model presented here, if the
accretion rate, at which the neutronization transition takes place, changes
during the evolution of the burst. The accretion rate at the transition decreases, for example, 
with increasing spin of the black hole (CB07). Since the black hole is expected to be
substantially span up because of accretion of matter during the evolution of
the burst (e.g. MacFadyen \& Woosley 1999), there is the possibility, though 
speculative at this level, that this leads to ejection of shells with varying
baryon loading.

\subsection{Photospheric emission}

Photospheric models for the origin of the prompt emission have been recently
explored for both fireballs (M{\'e}sz{\'a}ros \& Rees 2000; Ryde 2004;
Rees \& M{\'e}sz{\'a}ros 2005; Pe'er et al. 2006) and Poynting-flux dominated 
flows (Giannios 2006; Giannios \& Spruit 2007; hereafter GS07). Here, I focus mainly
to the photosphere of a Poynting-flux dominated flow since it is directly 
applicable to this work. 

In the photospheric model, the observed variability of the prompt emission is
direct manifestation of the central engine activity. Modulations of the 
luminosity and baryon loading of the GRB flow result in modulations of the location of the 
photosphere of the flow and of the strength and the spectrum of the
photospheric emission (Giannios 2006; GS07).
In particular, in GS07 it is demonstrated that {\it if} the increase of the
luminosity of the flow is accompanied by decrease of the baryon loading
such that\footnote{In GS07, the parameterization of the 
baryon loading of the flow is done by the magnetization $\sigma_0$ that is related to
$\eta$ through $\eta=\sigma_0^{3/2}$.} $\eta\propto L^{0.6}$, the photospheric
model can explain the observed narrowing of the width of the GRB pulses with increasing 
photon energy reported by Fenimore et al. (1995). The same $\eta$-$L$ scaling
also leads to the photospheric luminosity scaling with the peak of the
$\nu\cdot f(\nu)$ spectrum as $L_{\rm ph}\propto E_{\rm p}^{2}$ {\it during} the
burst evolution in agreement with observations (Liang et al. 2004).

The simple model for the connection
of the GRB flow to the properties of the central engine presented here  predicts that $L\propto {\dot
M}^{2.3...3.2}$ and  $\eta \propto {\dot M}^{1.3...2.2}$ during the
neutronization transition. The range in the exponents comes from the different assumptions on the disk-magnetic
field connection (see Sect. 3). This translates to $\eta\propto L^{0.6...0.7}$ which is very close that
assumed by GS07 to explain the observed spectral and temporal properties of the GRB light curves. 

Although the launched flow is Poynting-flux dominated,  
it is conceivable that it undergoes an initial phase of rapid
magnetic dissipation resulting to a fireball. The photospheric luminosity 
and the observed temperature of fireballs scale as $L_{\rm ph}\propto \eta^{8/3}L^{1/3}$,
$T_{\rm obs}\propto \eta^{8/3}L^{-5/12}$ respectively 
(M{\'e}sz{\'a}ros \& Rees 2000). Using the scaling $\eta\propto
L^{0.6...0.7}$ found in this work and identifying the peak of the
photospheric component with the peak of the emission $E_{\rm p}$ one finds that
$L_{\rm ph}\propto L^{1.9...2.2}$ and $E_{\rm p}\propto L^{1.2...1.4}$. The last scalings 
suggest that the photospheric emission from a fireball can further enhance variations in the
gamma-ray luminosity while $L_{\rm ph}$ and $E_{\rm p}$ follow the Liang et al. relation.
Still dissipative processes have to be considered in the fireball so that to explain the
observed non-thermal spectra. 

\section{Conclusions}

In this work, a mechanism is proposed by which moderate changes of the accretion rate 
at around $\dot{M}\sim 0.1$ $\Msun$/s to a few solar mass black hole can give  
powerful energy release episodes to the GRB flow. 
This mechanism is directly applicable to the collapsar scenario for GRBs
(Woosley; MacFadyen \& Woosley 1999) and can explain how moderate 
changes in the accretion rate result in extremely variable GRB light curves.

This mechanism operates when the inner part of the accretion disk makes the transition 
from advection dominance to neutrino cooling. This, rather sharp, transition is 
accompanied by steep increase of the density and the pressure in the disk 
(CB07; KM07). This leads to substantial increase of the magnetic field 
strength in the vicinity of the black hole and consequently boosts the Poynting 
luminosity of the GRB flow by a factor of $\sim 7-15$. At the same time, assuming that the 
ejection rate of material scales linearly with the accretion rate, the baryon
loading of the flow {\it decreases} by a factor $\sim 3-6$.
This results in a luminosity-baryon loading anticorrelation. 

The changes of the characteristics of the GRB flow can be
directly observed as modulations of the photospheric emission giving birth to
pulses with spectral and temporal properties similar to the observed ones
(GS07). The photospheric interpretation of the prompt emission is in agreement  
with the observed narrowing of the pulses with increasing photon energy (Fenimore et al. 1995)
and the luminosity-peak energy correlation during the evolution of GRBs (Liang et
al 2004). The Amati relation (Amati et al. 2002) is possibly result of the fact that more 
luminous bursts are on average less baryon loaded.  

\begin{acknowledgements}

I wish to thank H. Spruit for illuminating discussions on the
disk-magnetic-field coupling. 

\end{acknowledgements}

\end{document}